# Adapting sample size in particle filters through KLD-resampling

T. Li, S. Sun and T. Sattar

This letter provides an adaptive resampling method. It determines the number of particles to resample so that the Kullback–Leibler distance (KLD) between distribution of particles before resampling and after resampling does not exceed a pre-specified error bound. The basis of the method is the same as Fox's KLD-sampling but implemented differently. The KLD-sampling assumes that samples are coming from the true posterior distribution and ignores any mismatch between the true and the proposal distribution. In contrast, we incorporate the KLD measure into the resampling in which the distribution of interest is just the posterior distribution. That is to say, for sample size adjustment, it is more theoretically rigorous and practically flexible to measure the fit of the distribution represented by weighted particles based on KLD during resampling than in sampling. Simulations of target tracking demonstrate the efficiency of our method.

*Introduction:* The particle filter (PF) has been widely applied for nonlinear filtering due to its ability to carry multiple hypotheses relaxing the linearity and Gaussian assumptions. However, there are still challenges for PFs, such as specification of the sample size. Most existing particle filters use a fixed number of samples. However, since the complexity of the posterior distribution can vary drastically over time, the sample size should adjust online according to requirements of the system. One primary challenge in the application of particle filters is the design of an efficient method for sample size adjustment [1, 2].

The KLD provides a means to measure the fit of the distribution represented by weighted particles. It is used to determine the minimum number of particles needed to maintain the approximation quality in the sampling process, namely the KLD-sampling [3]. However, the samples are assumed to be coming from the true posterior distribution, which is actually not true. As an alternative, this letter applies the KLD measure in the resampling process thereby the distribution of interest is just the posterior distribution. Our approach provides a similar ability to adjust the sample size as the KLD-sampling and is more theoretically rigorous and practically flexible.

*KLD-sampling:* The KLD (also known as relative entropy) between the proposal ($q$) and true ($p$) distributions is defined in discrete form as

$$d_{KL}(p\|q) \triangleq \sum_x p(x)\log\left(\frac{p(x)}{q(x)}\right) = \sum_x W(x)q(x)\log W(x) \quad (1)$$

where $W(x) = p(x)/q(x)$.

It is derived in [3] to determine the required number $N$ of samples so that, with probability $1-\delta$, the Kullback–Leibler distance between sample-based maximum likelihood estimate (MLE) and the distribution of interest is less than a pre-specified error bound threshold $\varepsilon$, i.e.

$$N = \frac{1}{2\varepsilon}\chi^2_{k-1,1-\delta} \quad (2)$$

where $k$ is the number of bins with support, the quantises of the chi-square distribution is defined as

$$P(\chi^2_{k-1} \leq \chi^2_{k-1,1-\delta}) = 1-\delta \quad (3)$$

To save computation ((3) needs to be re-calculated online whenever a new particle is sampled), $\chi^2_{k-1,1-\delta}$ can be approximated by the Wilson-Hilferty transformation, which yields

$$N = \frac{k-1}{2\varepsilon}\left(1 - \frac{2}{9(k-1)} + \sqrt{\frac{2}{9(k-1)}}z_{1-\delta}\right)^3, \quad (4)$$

where $z_{1-\delta}$ is the upper quartile of the standard normal distribution. For typical values of $\delta$, the values of $z_{1-\delta}$ are available in standard tables.

This result gives the sample size needed to approximate a discrete distribution with an upper bound $\varepsilon$ on the KL-distance. It is incorporated into the sampling process namely the KLD sampling [3], in which the predictive belief state is used as the estimate of the underlying posterior. This, however, is actually not the case of particle filters where the samples come from the proposal distribution. As a result of this, the output of the KLD-sampling approach is based on statistical bounds of the approximation quality of samples that are actually drawn from the proposal distribution rather than the true posterior distribution. The mismatch between the true and the proposal distribution is ignored, see also [4, 5]. To avoid the mismatch, we apply the result of (4) in the resampling process to determine the total number of particles to resample. That is, we divide the particles (of the posterior distribution) into bins and count the number $k$ of bins in which at least one particle is resampled to determine the total number of particles to resample. Compared with the KLD-sampling, our approach (referred to as KLD-resampling) applies the result given in (4) to adjust the sample size in the resampling process rather than in the sampling process.

*Our approach:* There are two parts to the KLD-resampling approach which is described as in algorithm 1. One is to resample particles according to their weights one by one (individually and independently) until the required sample size (4) is satisfied. In the other part, the number $k$ of bins with support (in which at least one particle is resampled) and (4) are updated every time a new particle is (re)sampled. Except the resampling step, the other parts of the PF, i.e. the sequential importance sampling framework, do not need to change.

Similar to the KLD-sampling, the particles are sampled one by one individually until the required amount is achieved that is determined based on the KLD measure of the fit of the underlying distribution of particles. The advantage of our approach over the KLD-sampling is that the underlying distribution of our approach is just the posterior distribution while in the latter it is the predictive belief. The disadvantage of the KLD measure is that the particles need to be divided into bins in their state space, which can be highly inefficient when the state is of high dimension. For simplicity, we propose to divide the primary dimensions only, as proposed in [6]. In this case, the bin will be of low dimensionalities, see e.g. the simulation below.

---

**Algorithm 1: KLD-resampling**
**Inputs:** bound $\varepsilon$ and $\delta$, bin size, maximum number of samples $N_{max}$

**Initialization**: $k=0$; $i=0$; $N=1$; all bins are zero-resampled;
**while** ($i \leq N$ and $i \leq N_{max}$) **do**
    Randomly select one particle from the underlying particle set according to the weight (e.g. Multinomial resampling): $i:=i+1$
    **if** (the new resampled particle comes from non-resampled bin $b$) **do**
        Update the number of resampled bin: $k:=k+1$
        $b := $ resampled
        Update the required number $N$ of particle by (4)
    **end**
**end**

---

*Simulation:* The efficiency of the KLD measure for sample size adjustment for the particle filter has been demonstrated in the application of, but not limited to, mobile robot localization see [3, 4, 5]. For the sake of evaluating the sample size adjusting ability of our approach, we study the benchmark model of maneuvering target tracking. The target moves in the 2-dimensional plane according to a second order state space model

$$x_t = \begin{bmatrix} 1 & T & 0 & 0 \\ 0 & 1 & 0 & 0 \\ 0 & 0 & 1 & T \\ 0 & 0 & 0 & 1 \end{bmatrix} x_{t-1} + \begin{bmatrix} T^2/2 & 0 \\ T & 0 \\ 0 & T^2/2 \\ 0 & T \end{bmatrix} \begin{bmatrix} v_{1,t} \\ v_{2,t} \end{bmatrix} \quad (5)$$

where $x_t = [x_{1,t}, x_{2,t}, x_{3,t}, x_{4,t}]^T$, $[x_{1,t}, x_{3,t}]^T$ is the $x$-$y$ position while $[x_{2,t}, x_{4,t}]^T$ is the velocity at time $t$ and the sampling period T=1. The process noise $\{v_{1,t}\}$, $\{v_{2,t}\}$ are mutually independent zero-mean Gaussian white noise with respective standard deviation $\sigma_{v1}=0.001$ and $\sigma_{v2}=0.001$. The bearing-only measurement for an observer at the origin is given by

$$\theta_t = \arctan(x_{1,t}/x_{3,t}) + w_t \quad (6)$$

where $w_t$ is zero-mean Gaussian white noise with the standard deviation $\sigma_{w1}$=0.005.

The initial state of the target is $x_0 = [-0.05, 0.001, 0.7, -0.055]^T$, and prior to the measurements the state mean and standard deviations are assumed to be $x_0 = [0.0, 0.0, 0.4, -0.05]^T$, $\sigma_1$=0.5, $\sigma_2$=0.005, $\sigma_3$=0.3, $\sigma_4$=0.01. To note, the parameters of bound $\varepsilon$ and $\delta$, and especially the bin size are ad-hoc, and in practice it is necessary to carefully choose them to obtain the desired results. As suggested in [3], we use bound parameters $e$=0.15, $\delta$=0.01. More importantly, to save computation we divide the bins only in the 2-D position space and the bin size is chosen by considering of the system noise variance. We choose the bin size as the smaller of the standard deviations of the dynamics and the measurement, i.e. [0.001, 0.001]. The starting sample size is 1000 for all filters and the maximum sample size for KLD adaptive mechanism is 2000. The tracking scene and the sample size variation in one trial are given in Figure 1. The mean tracking position error and sample size of each filter of total 1000 trials are given in Figure 2. The tracking position error is the Euler distance between the estimate and the true position of the target which is defined as

$$error \triangleq \sqrt{(x_{1,t} - \hat{x}_{1,t})^2 + (x_{3,t} - \hat{x}_{3,t})^2} \qquad (7)$$

where $[\hat{x}_{1,t}, \hat{x}_{3,t}]^T$ is the $x$-$y$ position estimate of the target at time $t$.

The results show that the KLD-resampling PF obtains quite close estimation accuracy to the KLD-sampling PF and the basic particle filter. The KLD-resampling method can adjust the sample size efficiently as the KLD-sampling approach, as shown that when the estimation quality is reduced, more particles are generated. We conjecture that similar results will be obtained by the KLD-resampling in the context of mobile robot localization.

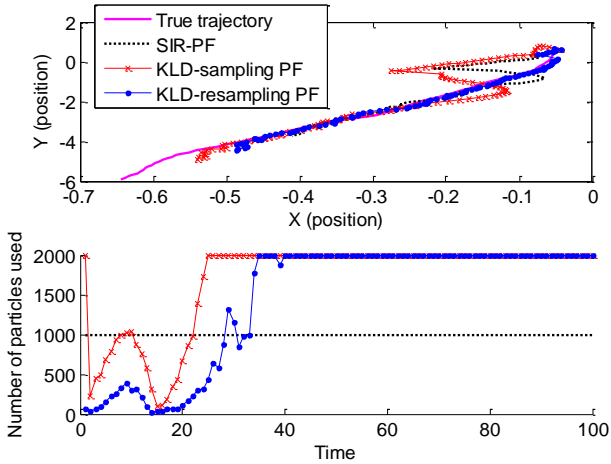

**Fig. 1** *The target tracking scene and sample size variation in one trial*

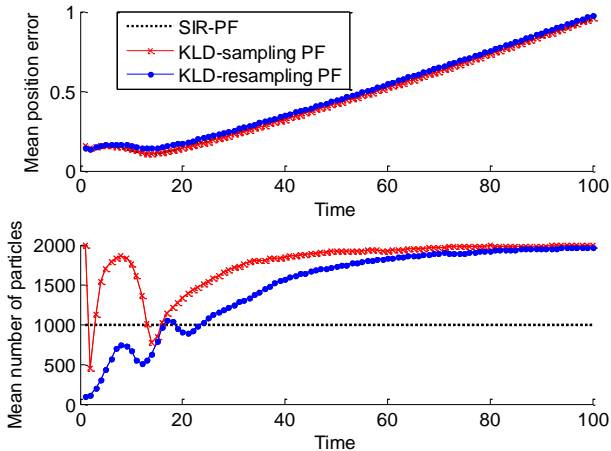

**Fig. 2** *The mean tracking error and sample size of 1000 trials*

*Conclusion:* An adaptive resampling method called KLD-resampling is proposed that determines the number of particles to resample based on the Kullback–Leibler measure of the fit of the posterior distribution represented by weighted particles. Our approach is based on the same theoretical ground as Fox's KLD-sampling but is more theoretically rigorous and practically flexible. The simulation in the context of maneuvering target tracking has demonstrated that our approach can efficiently adjust the sample size.

*Acknowledgments:* This work was supported by the National Natural Science Foundation of China (Grant No. 51075337).

Tiancheng Li and Shudong Sun (*The School of Mechatronics, Northwestern Polytechnical University, Xi'an, 710072, China.*)

E-mail: lit3@lsbu.ac.uk

Tariq P. Sattar (*London South Bank University, London, SE1 0AA, UK*)
T. Li is also with London South Bank University.